\documentclass[twocolumn]{aastex631}

\usepackage[utf8]{inputenc}
\usepackage{graphicx}
\usepackage{indentfirst}
\usepackage{amsmath}
\usepackage{natbib}
\setcitestyle{aysep={}}
\usepackage{rotating}
\usepackage{appendix}

\newcommand{\ha}{H$\alpha$}
\newcommand{\hb}{H$\beta$}
\newcommand{\mgb}{Mg {\footnotesize I}\textit{b}}
\newcommand{\CaII}{Ca {\footnotesize II}} 
 
\newcommand{\HII}{H{\footnotesize II}}

\newcommand{\NII}{[N~{\footnotesize II}]}

\newcommand{\OIII}{[O~{\footnotesize III}]} 
\newcommand{\SII}{[S~{\footnotesize II}]} 
\newcommand{\mbh}{$\mathrm{M}_{\mathrm{BH}}$}

\newcommand{\mgal}{$\mathrm{M}_{\mathrm{*}}$}

\newcommand{\sigstar}{$\sigma_{*}$}
\newcommand{\msun}{$\mathrm{M}_{\odot}$}

\begin{document}

\title{Magellan Spectroscopy of AGNs in Low-mass Galaxies: Scaling Relations and a Triple-Peaked AGN}

\author[0000-0003-1055-1888]{Megan R. Sturm}
\affil{Department of Physics, Montana State University, Bozeman, MT 59717, USA}

\author[0000-0001-7158-614X]{Amy E. Reines}
\affil{Department of Physics, Montana State University, Bozeman, MT 59717, USA}

\author[0000-0002-5612-3427]{Jenny Greene}
\affil{Department of Astrophysical Sciences, Princeton University, Princeton, New Jersey 08544,
USA}

\author[0000-0002-4587-1905]{Sheyda Salehirad}
\affil{Department of Physics, Montana State University, Bozeman, MT 59717, USA}


\begin{abstract}    

In this work, we aim to further populate the low-mass regime of black hole (BH) scaling relations to better understand the formation and growth mechanisms of central supermassive BHs. We target six galaxies that have been previously identified as hosting active galactic nuclei (AGN) based on optical spectroscopy from the Galaxy and Mass Assembly (GAMA) survey or the Sloan Digital Sky Survey (SDSS) with stellar masses {reported to be} $\mathrm{M_\star} < 5 \times 10^{9}$ \msun. Using follow-up optical spectroscopy from the Magellan Echellette Spectrograph (MagE), we extract galaxy velocity dispersions ($\sigma_\star$) and estimate virial BH masses from broad H$\alpha$ emission. We find that the galaxies in our sample do not deviate significantly from either the \mbh$-$\sigstar \: or \mbh$-$\mgal \:  scaling relations defined by higher mass galaxies. {Additionally, we identify one galaxy with triple-peaked \SII, \NII, \ha \: and \hb \: emission lines. This spectral shape is not shared by \OIII \: and, in fact, the \OIII \: line appears to have distinct kinematics from the other emission lines. Incorporating the spatial distribution of the various emission lines, we find that the galaxy spectrum is consistent with a prominent central AGN driving an outflow, surrounded by an extended ring/disk of gas predominantly ionized by shocks and/or star formation.} This work has implications for the demographics of BHs in low-mass galaxies and the role of AGN feedback.

\end{abstract}



\section{Introduction}
\label{sec:introduction}

Galaxy properties including velocity dispersion \citep{Gebhardt2000, Merritt2001}, bulge luminosity \citep{Marconi2003} and bulge mass \citep{Haring2004, McConnell2013, Kormendy2013} have been observed to correlate with the mass of their central supermassive black hole (SMBH), which are quantified with scaling relations. Historically, we have looked towards the centers of massive galaxies for dynamical BH mass measurements, or bright, easily observable broad-line AGNs and, as a result, we have amply populated the upper mass range of these scaling relations. In this upper mass range, we observe relatively tight scaling relations correlating increased BH mass with increased bulge mass, bulge luminosity and velocity dispersion.

While filling in the high-mass range of scaling relations is important, the low-mass regime is of particular interest in the context of BH growth and evolution. Studying the demographics of BHs at low galaxy/BH mass allows us to observe early phases of BH growth and place constraints on the initial seed properties/distinguish between possible seeding mechanisms. Both the slope and scatter in the low-mass regime of the \mbh $-$ \sigstar \: relation are expected to depend on the seed formation mechanism \citep{Volonteri2009}, with higher mass seeds resulting in a flattening of the relation around 10$^5$\msun. Furthermore, the BH fueling and growth efficiency are predicted to impact the distribution of galaxies along this relation \citep{Ricarte2018, Pacucci2018, Greene2020}.

\begin{deluxetable*}{ccccccc}
\tablecaption{Sample of Low-mass Active Galaxies}
\tablewidth{0pt}
\tablehead{
\colhead{ID} & \colhead{RA} & \colhead{DEC}  & \colhead{$z$} & \colhead{log ($\mathrm{M_\star}/$\msun)} & \colhead{BPT} & \colhead{{Broad \ha}} \\
\colhead{(1)} & \colhead{(2)} & \colhead{(3)} & \colhead{(4)} & \colhead{(5)} & \colhead{(6)} & \colhead{(7)} }
\startdata
 GAMA 2258819 &  30.87 & -9.49 & 0.252 & 9.60 &  AGN & {detected} \\
 GAMA 2123678 &  34.81 & -5.72 & 0.054 &  9.48 &  Comp & {detected} \\
 GAMA 5227891 &  341.86 &  -32.25 &  0.215 & 10.97\tablenotemark{\footnotesize{a}} & AGN & {hint} \\
 GAMA 5240292 &  343.27 & -33.98 & 0.224 &  9.59 &  AGN & {detected} \\
 GAMA 5275222 &  345.36 & -34.21 & 0.210 &  9.64 & AGN & {detected}\\
 SDSS J1605+0850 & 241.44 & 8.85 & 0.015 & 9.45 &  Comp & - \\
\enddata
\tablecomments{
Column 1: Identification number in GAMA/SDSS.
Column 2: RA {in degrees}.
Column 3: DEC {in degrees}.
Column 4: Redshift.
Column 5: {Log galaxy stellar mass as reported in \citet{Salehirad2022}.}
Column 6: BPT classification in the \OIII/H$\beta$ versus \NII/H$\alpha$ diagram from \citet{Reines2013} or \citet{Salehirad2022}. 
Column 7: {Whether broad \ha \: was detected (`detected'), not detected (`-') or if there was a hint of broad \ha \: (`hint') in either \citet{Salehirad2022} or \citet{Reines2013}.}
\tablenotetext{a}{{The total stellar mass for this object comes from the \texttt{ProPpectv03} table from GAMA (see Section \ref{sec:sample}).}}
}
\label{tab:sample}
\end{deluxetable*}

Directly observing BH seeds at the epoch which they are expected to have formed is currently not possible. Moreover, studies looking for low-mass BHs in the early Universe are biased towards only the most luminous sources that can be detected at such high redshift \citep[{e.g.,}][]{Onoue2023, Kocevski2023, Yang2023, Matthee2023, Juodvzbalis2023, Maoiolino2023, Harikane2023}. Therefore, we instead opt for filling in scaling relations with dwarf galaxies from the local Universe. Compared to larger galaxies, dwarf galaxies generally have experienced fewer merger events that can erase initial BH seed properties \citep{Volonteri2008}. Furthermore, since the smallest known massive BHs reside in dwarf galaxies ($M_{\rm BH} \sim 10^{4-5}$ \msun), we can use them to place upper limits on BH seed masses \citep{Reines2022}.

Two important works investigating the \mbh $-$ \sigstar \: relation in the low-mass regime are \citet{Xiao2011} and \citet{Baldassare2020}. The sample analyzed in \citet{Xiao2011} consists of 76 AGNs with \mbh $\lesssim 10^{6.5}$ \msun\ selected from the \citet{Greene2007} catalog of low-mass broad-line AGNs in SDSS DR4. \citet{Baldassare2020} presents new observations of eight dwarf galaxies ($M_\star \lesssim 3 \times 10^9$ \msun) with broad-line AGNs that were identified in \citet{Reines2013} (based on SDSS DR8), plus seven other dwarf galaxies with existing velocity dispersion and BH mass measurements. In both of these works, they find that their low-mass samples are generally in good agreement with the scaling relations derived from higher mass galaxies. However, as explained in \citet{Baldassare2020}, the results in the dwarf regime, where differences in seeding models should become apparent in the scaling relations, are still limited by the sample size and possibly selection bias. {Since the Eddington limit of a BH is proportional to its mass and continuous accretion at the Eddington limit is unlikely \citep{Schulze2010}, observationally, we are biased towards finding higher mass BHs that are accreting at lower rates. This means that current samples of BHs occupying the low mass regime of scaling relations could just represent an upper bound on the true population demographics.}

In this work, we present new Magellan spectroscopy of six galaxies to obtain velocity dispersions and virial BH masses. Ultimately, we are able to place four of these objects on the \mbh $-$ \sigstar \: and \mbh $-$ \mgal \: scaling relations and compare to canonical relations. Additionally, we find that one galaxy has triple-peaked emission lines. Our paper is laid out as follows. We present our sample in Section \ref{sec:sample} and describe our new MagE observations and data reduction in Section \ref{sec:obs_red}. Our spectral analysis, including galaxy continuum/absorption line + AGN emission line fitting, is presented in Section \ref{sec:spec_analysis}. {Specifically in Section \ref{sec:triple_peaked}, we show a unique galaxy with triple-peaked emission lines and describe the spectral fitting for it. We discuss how our galaxies compare to canonical scaling relations in Section \ref{sec:scaling_relations}. Finally, we discuss possible physical interpretations of our triple-peaked emission line galaxy in Section \ref{sec:triple_peaked_interpretation}.} Throughout this work we adopt $\mathrm{H}_0 = 70$ km $\mathrm{s}^{-1}$  $\mathrm{Mpc}^{-1}$.

\begin{figure*}
    \centering
    \includegraphics[width=15cm]{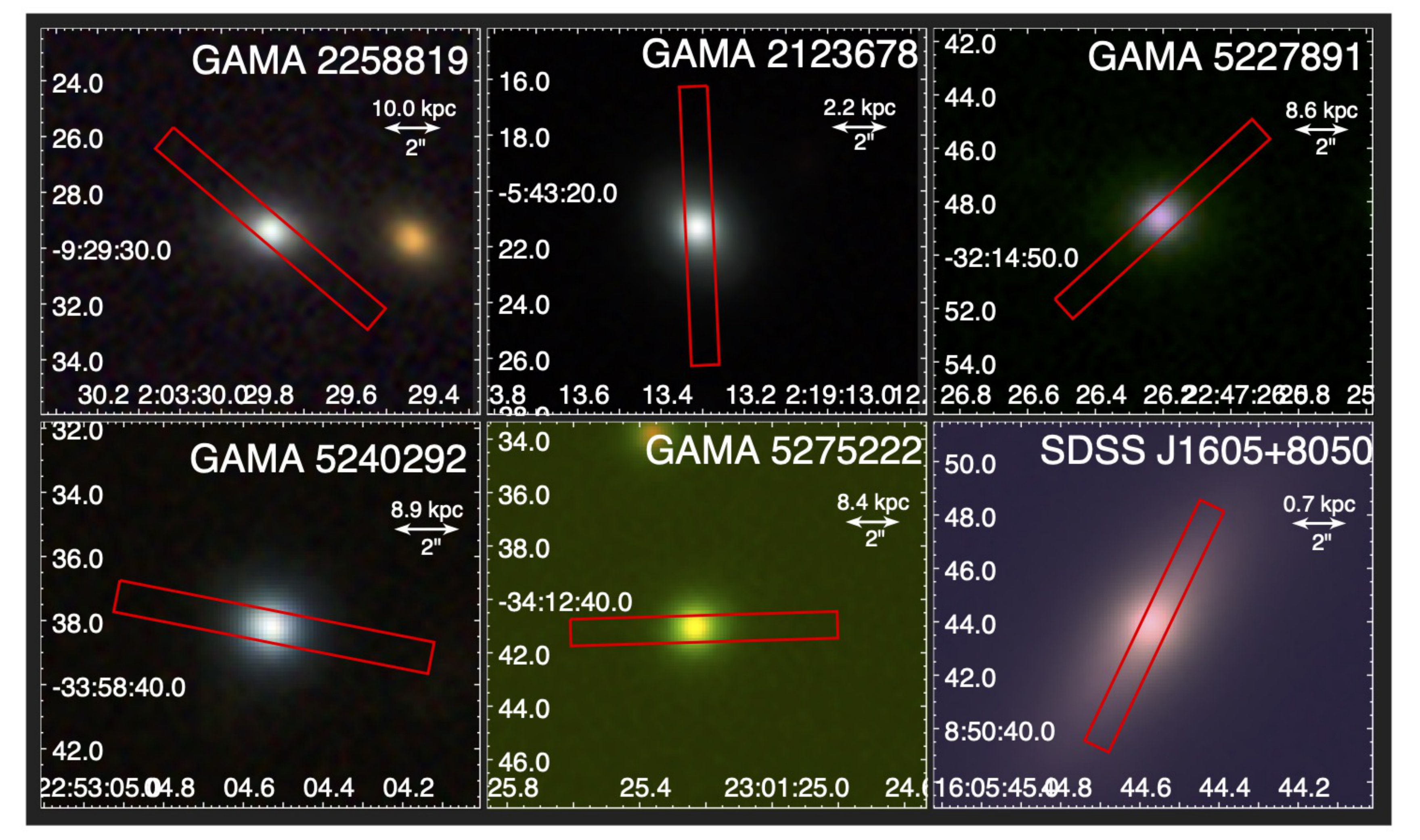}
    \caption{{DESI Legacy Imaging Survey SkyViewer Images of our six low-mass galaxies. {We overlay the 10" $\times$ 1" slit used to observe each target in red. We also provide a scale bar of length 2" along with the corresponding length in kpc. }}}
    \label{fig:galaxy_images}
\end{figure*}


\section{Sample of Low-Mass Active Galaxies}
\label{sec:sample}

Our targets primarily come from the new sample of AGNs in low-mass galaxies discovered in the Galaxy and Mass Assembly (GAMA) survey by \citet{Salehirad2022} using optical spectroscopy. They restrict the SED-derived stellar masses {(as reported from GAMA)} of their galaxies to $\mathrm{M_\star} \leqslant 10^{10}$ \msun. Additionally, following \citet{Reines2013}, they impose restrictions on the signal-to-noise (S/N) ratio and equivalent width (EW) of certain emission lines: \ha, \OIII \: and \NII \: must have S/N $\geqslant$ 3 and EW $>$ 1 \AA \: and \hb \: must have S/N $\geqslant$ 2. Finally, they make a redshift cut (z $\leqslant$ 0.3) to make sure the \SII \: doublet is within the observed wavelength range, leaving them with a final parent sample of 23,460 galaxies.

In their work, they use various emission line diagnostics to distinguish AGNs from star-forming galaxies and produce a final sample of 388 new low-mass, spectroscopic AGNs. We select a subset of five galaxies to further investigate in this work. Our galaxies were all classified as either AGNs or composites using the Baldwin, Philips and Terlevich \citep[BPT;][]{Baldwin1981} diagram. Moreover, four of the galaxies have broad \ha \: detected in the GAMA spectra and the remaining galaxy {(GAMA 5227891)} was identified as a narrow-line AGN displaying an asymmetric \ha \: line, hinting at a potential additional, underlying broad component that could be a solid detection with higher resolution spectra, similar to the case of RGG 118 \citep{Baldassare2015}. With the detection of broad \ha \: emission, we are able to derive BH mass estimates. Lastly, all of the galaxies we selected to have stellar masses of $\mathrm{M_\star}<5\times10^{9}$\msun \: {reported in \citet{Salehirad2022}}. {However, for one object (GAMA 5227891), the reported stellar mass was anomalously low and, upon checking the imaging for this galaxy, we found the $g$-band image is missing data at the location of the target.
Therefore, we instead use the total stellar mass presented in the GAMA \texttt{ProSpectv03} table \citep{Bellstedt2020} of $\mathrm{M_\star}=10^{10.97}$ \msun, {since ProSpect deals with missing data appropriately. This value is consistent with our estimate of the stellar mass using the $u$ and $i$-band images and the color-dependent mass-to-light ratios in \citet{Zibetti2009}.} While this mass is well above the dwarf galaxy regime, we retain this object for further study as its spectrum reveals a rare triple-peaked profile (\S \ref{sec:triple_peaked}).}


The final galaxy in our sample was selected as a filler object from the work of \citet{Reines2013}, where they selected AGNs in dwarf galaxies (as described above) based on optical spectra from SDSS. After imposing the S/N, EW and redshift cuts, their parent sample includes 25,974 galaxies. The galaxy we select is ID 128 in their work (ID J1605+0850 in SDSS), which was classified as a BPT composite galaxy and did not show signs of broad H$\alpha$ in the SDSS spectrum. Properties of our sample are provided in Table \ref{tab:sample}. {Images of our 6 target galaxies from the DESI Legacy Imaging Survey SkyViewer \citep{Dey2019} are shown in Figure \ref{fig:galaxy_images}}.

\section{Observations and Data Reduction}
\label{sec:obs_red}

We obtain new, high-resolution spectra of our galaxies using the Magellan Echellette Spectrograph (MagE) on the 6.5 m Clay Telescope at Las Campanas Observatory. All of our targets were observed on 22 August 2023 using a 1 arcsecond slit for three 30 minute exposures, totalling 1.5 hours of integration time each. In echellette mode, MagE spans the full rest-frame optical range from 3100--10000\AA \: and has a spectral resolution of R$\sim$4100 ($\sim 20$~km s$^{-1}$). 

We start our data reduction using the reduction pipeline \texttt{pypeit} \citep{pypeit:prochaska2020a, pypeit:prochaska2020b}. The pipeline performs bias subtraction, flat-fielding, wavelength calibration (using exposures of a ThAr lamp) and flux calibration. We use the standard star EG274 \citep{Hamuy1992} for the flux calibration. {The pipeline performs optimal extraction using the Horne Algorithm \citep{Horne1986}. This algorithm identifies the object in each 2-D spectrum and applies non-linear weights to each pixel in order to optimally enclose the object while minimizing the inclusion of noisy pixels on the edges.} The final product of this pipeline is a 1-D spectrum for each echellette order in units of $10^{-16}$ erg s$^{-1}$ cm$^{-2}$ \AA$^{-1}$. For the subsequent fitting, we combine all available orders into a complete 1-D spectra for each target.

{We note that the inclusion of a disk component, which is more likely relevant in low-mass galaxies, could affect the placement of a galaxy on the \mbh - \sigstar \: relation \citep{Woo2006, Bennert2011, Harris2012, Woo2013, Bennert2015}. For dwarf galaxies hosting AGNs with high resolution Hubble Space Telescope observations, \citet{Kimbrell2021} find bulge components to have effective radii of $\sim 0.1-1.6$ kpc, with a median of $\sim 0.3$ kpc. Therefore, we expect that we are fully encompassing the bulge component and getting some contribution from the disk of each galaxy as well (see Figure \ref{fig:galaxy_images}).}


\section{Spectral Analysis}
\label{sec:spec_analysis}

\subsection{Continuum and Absorption-Line Subtraction}
\label{sec:continuum}

\begin{figure*}
    \centering
    \includegraphics[width=18cm]{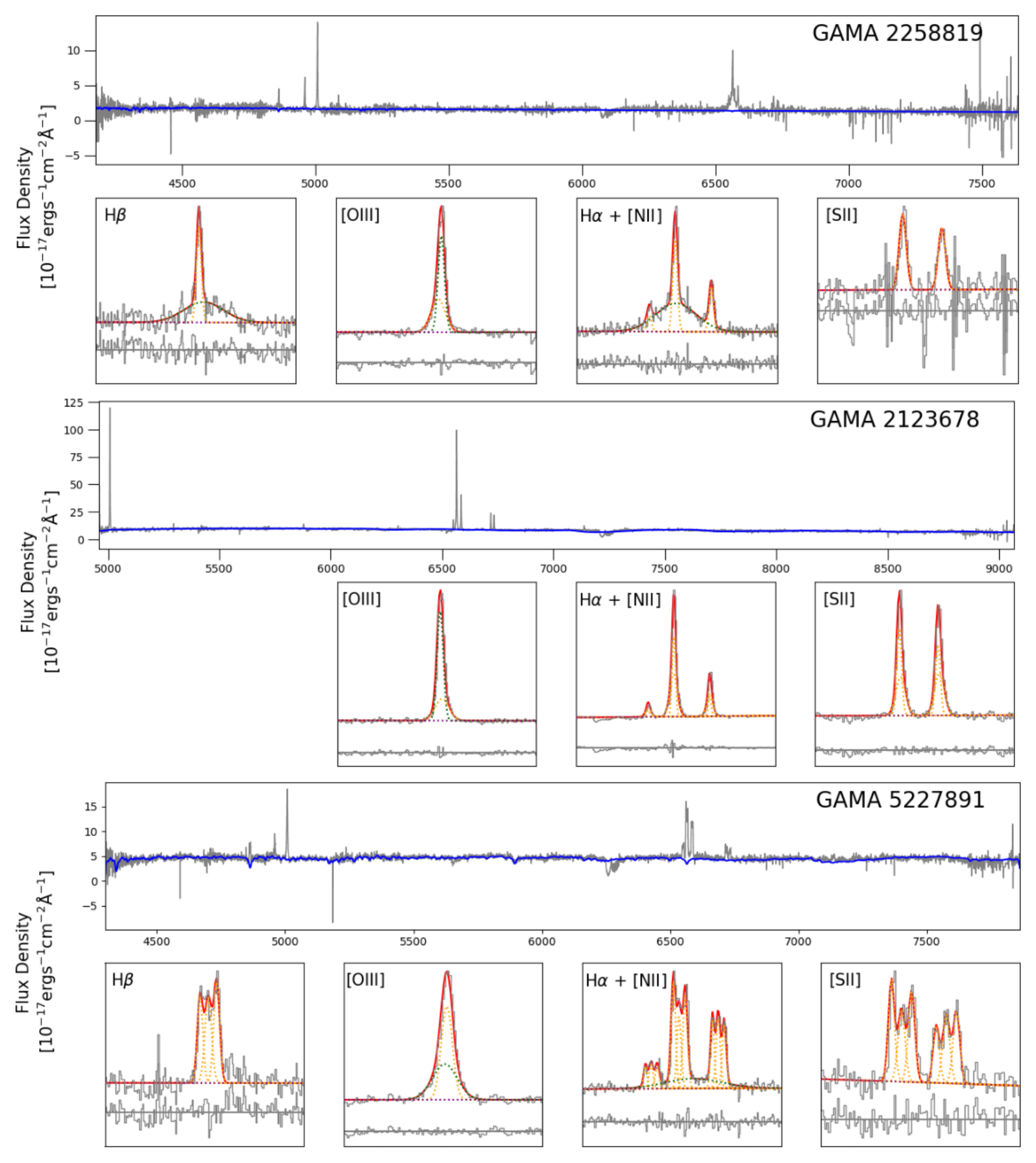}
    \caption{Absorption and emission line fits for three of our galaxies (the remaining three galaxies are shown in Figure \ref{fig:spec_all_2}). For each target, the top panel shows the full MagE galaxy spectrum with the \texttt{pPXF} fit to the stellar continuum overlaid in blue. The bottom panels show close-up views of relevant emission lines. The narrow line fits are shown in orange and the broad line fits (when present) are shown in green. We show the residuals below each emission line fit with a vertical offset for clarity. We perform a custom fitting process to the triple-peaked emission line galaxy GAMA 5227891 as described in Section \ref{sec:triple_peaked}.}
    \label{fig:spec_all_1}
\end{figure*}

\begin{figure*}
    \centering
    \includegraphics[width=18cm]{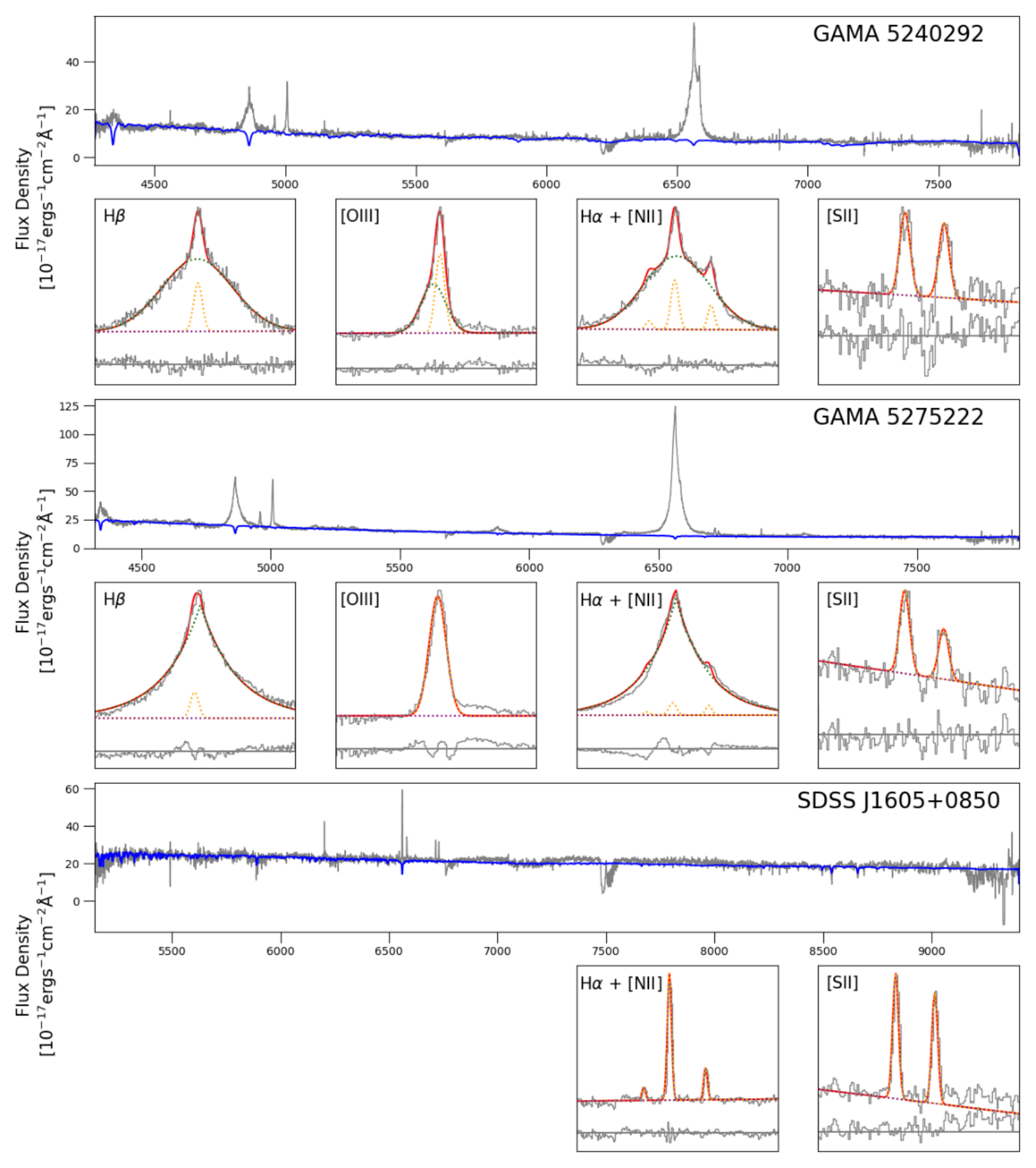}
    \caption{Same as Figure 1 but for the three remaining galaxies.  For GAMA 5275222, we model the broad component to the Balmer lines with an exponential profile. }
    \label{fig:spec_all_2}
\end{figure*}

Before we can fit an AGN emission line spectrum, we must first remove the host galaxy contribution from the observed spectrum. We fit the continuum and absorption lines in our galaxy spectra using the Penalized Pixel Fitting software \citep[\texttt{pPXF;}][]{Cappellari2004}. We first trim each galaxy spectrum to remove the edge orders with relatively high noise. 
We then fit the trimmed spectra using GALAXEV stellar population synthesis model templates of stellar continua and absorption lines \citep{Bruzual2003}. These templates have a FWHM of 3.0\AA \: in the wavelength range of our spectra. Additionally, we include multiplicative and additive polynomials of degree 1 and 2, respectively. This allows us to correct for higher order fluctuations and get a more accurate velocity dispersion measurement \citep{Cappellari2017}. We show these fits in Figures \ref{fig:spec_all_1} and \ref{fig:spec_all_2}. This best fit model is subtracted from each spectrum to obtain a pure AGN emission line spectrum to be used for emission line fitting.

\begin{deluxetable*}{lccccccccc}
\tablecaption{Emission Line Fluxes}
\tablewidth{0pt}
\tablehead{
\colhead{ID} & \colhead{H$\beta_{\mathrm{n}}$} & \colhead{H$\beta_{\mathrm{b}}$}  & \colhead{\OIII$_\mathrm{n}$} & \colhead{\OIII$_\mathrm{b}$} & \colhead{H$\alpha_{\mathrm{n}}$} & \colhead{H$\alpha_{\mathrm{b}}$} & \colhead{\NII} & \colhead{\SII} & \colhead{\SII} \\
\colhead{ } & \colhead{$\lambda$4861} & \colhead{$\lambda$4861} & \colhead{$\lambda$5007} & \colhead{$\lambda$5007} & \colhead{$\lambda$6563} & \colhead{$\lambda$6563} & \colhead{$\lambda$6583} & \colhead{$\lambda$6716} & \colhead{$\lambda$6731} \\
\colhead{(1)} & \colhead{(2)} & \colhead{(3)} & \colhead{(4)} & \colhead{(5)} & \colhead{(6)} & \colhead{(7)} & \colhead{(8)} & \colhead{(9)} & \colhead{(10)}
}
\startdata
GAMA 2258819 & 7 (1) & 11 (1) & 17 (2) & 23 (23) & 21 (1) & 57 (2) & 10 (1) & 6 (1) & 5 (1) \\
GAMA 2123678 &  &  & 99 (12) & 222 (13) & 309 (16) & \nodata & 106 (5) & 45 (4) & 42 (4) \\
GAMA 5227891 - central \tablenotemark{\footnotesize{a}} & 10 (5) &  \nodata & 36 (4) & 31 (3) & 21 (2) & 38 (4) & 20 (1) & 6 (1) & 5 (1) \\ 
GAMA 5227891 - red \tablenotemark{\footnotesize{a}} & 8 (3) &  \nodata &  \nodata & \nodata & 32 (1) &  \nodata & 17 (1) & 7 (1) & 4 (1) \\ 
GAMA 5227891 - blue \tablenotemark{\footnotesize{a}} & 10 (3) &  \nodata &  \nodata & \nodata & 29 (1) &  \nodata & 17 (1) & 7 (1) & 5 (1) \\ 
GAMA 5240292 & 41 (3) & 500 (8) & 47 (3) & 78 (78) & 110 (5) & 1402 (15) & 47 (3) & 16 (1) & 15 (1) \\
GAMA 5275222 & 41 (5) & 1265 (2) & 255 (4) & \nodata & 64 (10) & 3997 (6) & 41 (4) & 24 (1) & 14 (1) \\
SDSS J1605+0850 &  &  &  &  & 143 (2) & \nodata & 32 (2) & 32 (3) & 29 (3) \\
\enddata
\tablecomments{
Column 1: Identification number in GAMA/SDSS.
Column 2-10: Emission line flux (and flux uncertainties in parentheses) in units of 10$^{-17}$ erg s$^{-1}$ cm$^{-2}$ resulting from the \texttt{LMFIT} line fitting described in Section \ref{sec:emission}. A blank space indicates that the emission line is not present in the MagE spectrum for that object {due to the wavelength range of MagE and the redshift of the galaxy}. A three-dot ellipsis indicates that no line is detected. 
\tablenotetext{a}{The fitting process for this object is described in Section \ref{sec:triple_peaked}. The \OIII\ fluxes are reported for Model 2.} 
}
\label{tab:fluxes}
\end{deluxetable*}

\subsection{AGN Emission Line Fits}
\label{sec:emission}

We fit the relevant AGN emission lines in our host-galaxy-subtracted spectra using the least-squares fitting methods of \texttt{LMFIT} \citep{Newville2014}. All lines are modeled using a sum of various Gaussians + a linear component to account for any errors in the \texttt{pPXF} stellar continuum fitting. One of our galaxies displays triple-peaked emission lines and we will discuss it further in Section \ref{sec:triple_peaked}. For the remaining five galaxies, we follow the procedure of \citet{Reines2013}, described below.

We first fit the \SII \: doublet using both one and two Gaussian components for each line. We fix the separation of the two peaks of the doublet to be the rest-frame wavelength separation and constrain the widths of the two peaks to be equal in velocity space. For the two component model, we constrain the relative heights, widths, and positions of each component to be equal for both lines in the doublet. If adding the second components decreases the $\chi^2$ from the single component model by 20\% or more, we assume that a two component model is preferred. Only one of our galaxies (ID 2123678) required a two Gaussian component model. For the remaining four galaxies, a single Gaussian component is preferred for each line in the doublet.

Once we have the best-fit Gaussian parameters for the \SII \: lines, we use them as a template for the narrow H$\alpha$ and \NII \: emission lines. We fix the separation of the \NII \: lines to their rest-frame wavelength separation and require that $\sigma_{\NII}$ is equal to $\sigma_{\SII}$ in velocity space. We also constrain $\sigma_{\mathrm{H\alpha,narrow}}$ to be $<1.25 \times \sigma_{\SII}$ in velocity space. Additionally, if necessary, we include a broad H$\alpha$ component with no restrictions except that $\sigma_{\mathrm{H\alpha,broad}} > \sigma_{\mathrm{H\alpha,narrow}}$ \footnote{{Galaxy GAMA 5275222 has very broad tails on its Balmer emission lines. Following from \citet{Kollatschny2013}, we test fitting the broad lines using Gaussian (doppler motion), Lorentzian (turbulent motion), exponential (electron scattering) and logarithmic (inflow/outflow motion) profiles. The best model for the broad lines in this galaxy is an exponential profile \citep[e.g.,][]{Laor2006}.} }. The fit including a broad H$\alpha$ component is preferred if the {reduced} $\chi^2$ is improved by at least 20\%. Three out of the five galaxy spectra that we fit in this section are best fit with a model including broad H$\alpha$ emission, consistent with the findings from \cite{Salehirad2022}. One of the galaxies without detectable broad H$\alpha$ emission, SDSS J1605+0850, also did not show broad H$\alpha$ in the previous work done by \citet{Reines2013}. However, \citet{Salehirad2022} did detect weak broad H$\alpha$ in the spectrum of GAMA 2123678. This is likely because our MagE spectrum has higher spectral resolution than GAMA and so the second component in the \SII \: doublet was not detected in the GAMA spectrum.

We fit the H$\beta$ emission line in the same way as the H$\alpha$ line; constraining its width to be $<1.25 \times \sigma_{\SII}$ in velocity space and allowing for a broad component (with width larger than the narrow component) if it reduces the {reduced} $\chi^2$ value by $>$20\%. Since two of our galaxies (GAMA 2123678 and SDSS J1605+8050) are at such low redshift, the H$\beta$ emission line is outside of the wavelength range observed with MagE. Of the remaining three galaxies, all of them are consistent with having broad H$\beta$.

We fit the \OIII $\lambda$5007 line independently because the profile of this line does not necessarily resemble the other narrow emission lines \citep{Greene2005}.  \OIII \: sometimes displays a broad component which can be due to ionized gas outflows \citep{Heckman1984}. These outflows have been found to be generally blueshifted in AGNs \citep{ZakamskaGreene2014, salehirad2024}. Since the \OIII \: line is not produced in the broad-line region around a BH, additional velocity components in this line can be a tracer of ionized gas dynamics in the narrow-line region. Therefore, we use a Gaussian model with no constraints to fit this emission line. Once again, we opt for a model including a broad component if it improves the {reduced} $\chi^2$ value by at least 20\%. Due to its redshift, for one galaxy (SDSS J1605+8050) the \OIII\ emission line is outside of the wavelength range observed with MagE. For the remaining galaxies, one (GAMA 5275222) is best fit with a single narrow component and three are best fit with a narrow + broad component. These broad components are generally blue-shifted, ranging from 30-117 km s$^{-1}$ offset. One galaxy (ID 2123678) has a broad component that is red-shifted by 55 km s$^{-1}$. The widths of these broad components range from 55-220 km s$^{-1}$.

The best-fit emission line models for our galaxies are shown in Figures \ref{fig:spec_all_1} and \ref{fig:spec_all_2}. The fluxes we measure are reported in Table \ref{tab:fluxes}. For two of our galaxies (GAMA 2123678 and SDSS J1605+0850) not all of the emission lines needed for the BPT diagram are within the observed MagE wavelength range and so we cannot measure their narrow emission line ratios. The three remaining galaxies are found to be BPT AGNs, which is consistent with the previous findings of \citet{Salehirad2022}.

{The procedure described above involves selecting models with more components if adding those components decreases the reduced $\chi^2$ by 20\% or more. While this criterion is empirical, it has been used in multiple previous works to establish AGN samples \citep[e.g.,][]{Reines2013, Salehirad2022, Hao2005}. Furthermore, it seems to select reasonable models for our sample based on visual inspection. As an additional check that this 20\% cut on reduced $\chi^2$ improvement is sufficient, we compare the Akaike Information Criteria (AIC) between single and multi-component models and find that the results are broadly consistent with the method above. Therefore, for consistency with previous works and since the results agree with using the AIC in all cases except when the AIC values are close, we continue to select our models based on the reduced $\chi^2$ value.}


\subsection{BH Mass Estimates}
\label{sec:BH_masses}

{In agreement with previous work, three galaxies with detected broad \ha \: in \citet{Salehirad2022} also display broad \ha \: in their MagE spectra\footnote{{The one galaxy with a `hint' of broad \ha \: in  \citet{Salehirad2022} will be discussed in Section \ref{sec:triple_peaked}}.}}. {While transient broad H$\alpha$ in star-forming galaxies can be attributed to supernovae, persistent broad H$\alpha$ with BPT AGN narrow emission line ratios provide a robust indicator of the presence of an AGN \citep{Baldassare2016}.} For these galaxies, we are able to estimate their BH masses. Assuming the gas in the broad-line region (BLR) is virialized, the BH mass scales proportional to the size of the BLR and the {average gas velocity squared}. {We measure the average gas velocity as the full width at half maximum (FWHM) of the broad \ha \: line and estimate the size of the BLR using the broad \ha \: luminosity as a proxy \citep{Greene2005}. } Therefore, as presented in \citet{Reines2013}, the virial BH mass can be estimated as:

\begin{equation}
    \begin{split}
        \mathrm{log \left( \frac{M_{BH}}{M_\odot} \right) = log \epsilon + 6.57 + 0.47 log \left( \frac{L_{H\alpha}}{10^{42} erg s^{-1}} \right) } \\
        \mathrm{+ 2.06 log \left( \frac{FWHM_{H\alpha}}{10^3 km s^{-1}} \right)}
    \end{split}
\end{equation}

\noindent
We use $\epsilon=$1 as in \citet{Reines2013}. Uncertainties associated with using this method are $\sim$0.5 dex.

\begin{figure*}[!t]
    \centering
    \includegraphics[width=18cm]{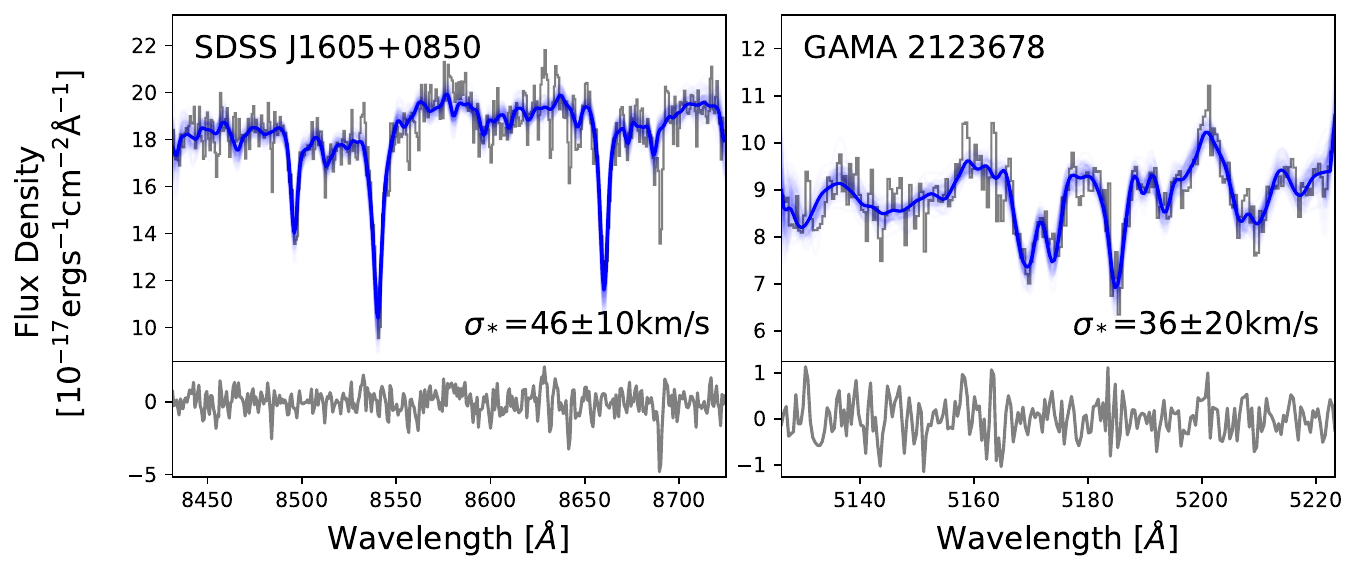}
    \caption{\texttt{pPXF} fits for the two galaxies in our sample exhibiting strong absorption lines (top) and the resulting residuals (bottom). The resulting fits from all {100} noise realizations from the Monte Carlo bootstrapping are shown in blue. }
    \label{fig:pPXF_fits}
\end{figure*}

{The $(L_{H\alpha})_b$ range from $10^{39.8-42.6}$ erg s$^{-1}$ and the broad H$\alpha$ FWHMs range from 1167-2030 km s$^{-1}$. The \mbh \: we calculate using these luminosities and FWHMs range from $10^{6.2-7.3}$ \msun.} Our mass estimates are presented in Table \ref{tab:results}. These values are consistent with those previously calculated using the GAMA spectra in \citet{Salehirad2022}, with our BH masses differing by 0.15 dex on average (for these four galaxies).

{Two of our galaxies (GAMA ID 2123678 and SDSS J1605+8050) do not have detectable broad \ha \: and so we cannot derive a virial BH mass for these objects. However, following the analysis done in \citet{Reines2013}, we can estimate an upper limit on the BH mass based on the minimum observable broad \ha \: flux. We estimate this by adding a Gaussian to the \ha \: line with FWHM 500 km s$^{-1}$ and incrementally increase the height until the broad component is detected by our code. The minimum detectable broad \ha \: luminosities for GAMA 2123678 and SDSS J1605+8050 are 1 $\times 10^{39}$ erg s$^{-1}$ and 2 $\times 10^{38}$ erg s$^{-1}$, which translate to upper BH mass limits of  $10^{4.58}$ \msun and $10^{4.22}$ \msun, respectively. Using the \citet{Reines2015} AGN \mbh $-$ \mgal \: relation for their stellar masses results in BH mass estimates of \mbh$\sim10^{5.8}$ \msun\ for both galaxies. Using the \citet{Kormendy2013} \mbh $-$ \sigstar \: relation for their velocity dispersions (see \ref{sec:abs_sigma}) results in BH mass estimates of \mbh$\sim10^{5.2}$ \msun \: for GAMA 2123678 and \mbh$\sim10^{5.7}$ \msun \: for SDSS J1605+8050. Both of these scaling relations give roughly consistent BH mass estimates. These measurements are not consistent with the upper mass limit derived using the minimum detectable broad \ha \: luminosity, however, we note that limit was calculated assuming the AGN has a broad-line region, which may not be the case for low-luminosity AGNs  \citep[e.g., ][]{Tran2001,Gu2002,Tran2003, Elitzur2009}.} {That calculation would also not hold for the case of Type II AGNs}.

\subsection{Stellar Velocity Dispersion Measurements from \mgb \: and \CaII \: } 
\label{sec:abs_sigma}

In order to obtain more precise velocity dispersion measurements, we trim each spectrum around both the \mgb \: (5167 \AA, 5172 \AA \: and 5183 \AA) and \CaII \:  triplets (8498 \AA, 8542 \AA \: and 8662 \AA) and fit the spectrum again using \texttt{pPXF}. Since we have focused into the specific absorption line regions, we fit each region using stellar spectra rather than the model stellar population templates. For the \mgb \: region, we follow the work of \citet{Baldassare2016} and fit the trimmed spectra using a set of a wide range of stars from the ELODIE spectral database \citep{Prugniel2001}. These spectra span 3900-6800\AA \: and have spectral resolution R = 10,000. For the \CaII \: region, we follow \citet{Caglar2020} and use the G, M and K stars in the X-shooter spectral library from \citet{Chen2014}. These spectra span 3000-10200\AA \: and have resolution R=10,000. We also include multiplicative and additive polynomial of degrees 3 and 12, respectively \footnote{{Increasing the polynomial degrees aids in obtaining a more accurate fit by correcting for spectral calibration inaccuracies, dust reddening, and sky subtraction errors. Specifically the additive polynomials can help to deepen individual absorption lines \citep{Cappellari2017}}.}.

We calculate uncertainties associated with this fitting process via a Monte Carlo bootstrapping method; iteratively adding noise to each flux point in the subsequent spectra based on a Gaussian with $\sigma$ equal to the standard deviation of the residuals from the previous fit. We do this for {100} noise realizations, where {$\sigma_* > \sigma_{\mathrm{instrument}}$;} if the resulting velocity dispersion found by the \texttt{ppxf} fit is smaller than our instrumental resolution, we discard the fit. For our final reported velocity dispersion we remove the effects of the instrumental dispersion from the $\sigma_*$ found by \texttt{pPXF}. 

Two of our galaxies display absorption lines in both the \CaII \:  and the \mgb \: triplets. For galaxy SDSS J1605+0850, the \CaII \:  triplet is significantly more {visually prominent} than the \mgb \: triplet, while for galaxy GAMA 2123678, the \mgb \: triplet is the more distinctive absorption line complex. Therefore, we take the stellar velocity dispersion measurement for the more distinctive complex. {SDSS J1605+0850 has a velocity dispersion of 46 $\pm$ 10 km s$^{-1}$ and GAMA 2123678 has a velocity dispersion of 36 $\pm$ 20 km s$^{-1}$.} The results of this fitting are shown in Table \ref{tab:results} and Figure \ref{fig:pPXF_fits}.

Four of our galaxies do not have detectable stellar absorption lines. For these galaxies, we estimate stellar velocity dispersion measurements using the ionized gas velocity dispersion as described in the following section (\S \ref{sec:nii_sigmas}).

\subsection{Velocity Dispersion Measurements from \NII}
\label{sec:nii_sigmas}

{Since the galaxy absorption lines were not detectable for multiple galaxies, we were not able to obtain stellar velocity dispersion measurements in the same way as described in Section \ref{sec:abs_sigma}}. {For these galaxies, we instead use the ionized gas velocity dispersion ($\sigma_g$), measured through the FWHM of the \NII \: lines, as a proxy \citep{Barth2008,Xiao2011,Baldassare2015}. }The scatter around the $\sigma_* - \sigma_g$ correlation is $\sim$0.15 dex so our \NII-derived velocity dispersion measurements carry extra contribution to the uncertainty than those directly measured through the \texttt{pPXF} fitting. The resulting $\sigma_*$ values for our galaxies {range from 37-80 km s$^{-1}$} and are shown in Table \ref{tab:results}. 

We note that for GAMA 2123678 and SDSS J1605+0850 we were able to obtain stellar velocity dispersion measurements using both stellar absorption lines and the \NII \: emission lines. The $\sigma_*$ values we calculate for both of these objects are consistent within their uncertainties for both methods.

\begin{figure*}
    \centering
    \includegraphics[width=16cm]{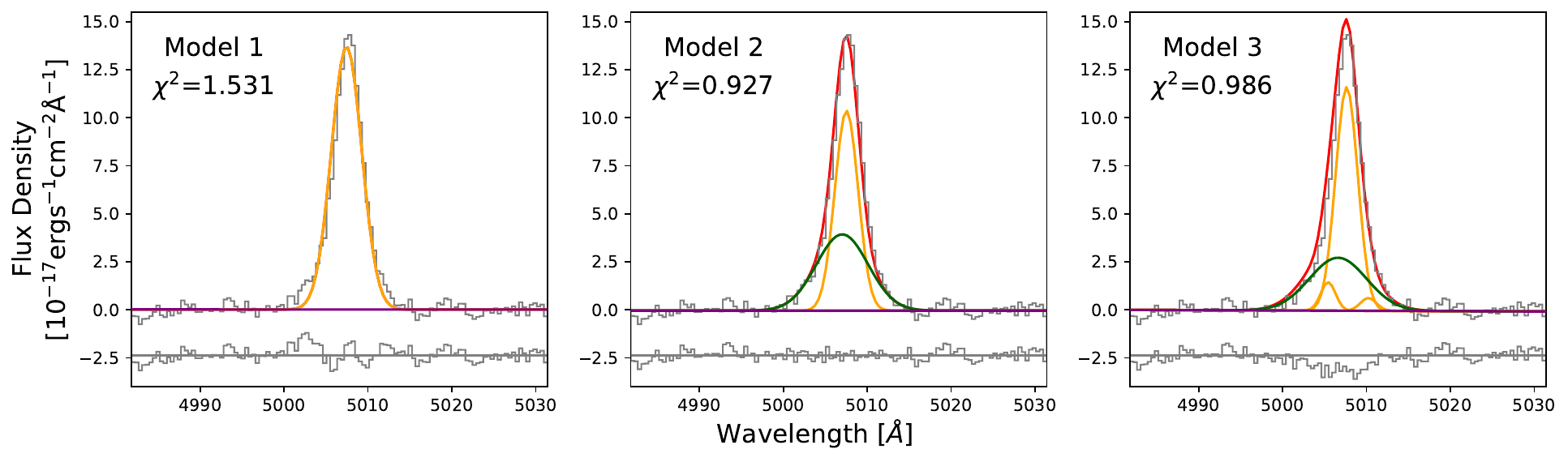}
    \caption{The three models we consider for fitting the \OIII \: emission line (see \S\ref{sec:triple_peaked}). Model 1 includes a single Gaussian component. Model 2 includes one narrow component + one broad component. Model 3 includes three narrow components + one broad component. For the red/blue peaks in Model 3, we constrain the widths/centers based on the corresponding peaks in the other emission lines. In all three models, narrow components are shown in yellow, broad components are shown in green and the full sum of all components is shown in red. {We show the residuals below each emission line fit with a vertical offset for clarity.} {The reduced $\chi^2$ for each model is given in the top left of each panel.}}
    \label{fig:oiii_fits}
\end{figure*}

\subsection{A Triple-Peaked Emission Line Galaxy}
\label{sec:triple_peaked}

One of our galaxies, GAMA 5227891, has triple-peaked emission lines in H$\alpha$, \NII, \SII \: and H$\beta$\footnote{We ensure the triple peaks are real, as described in the Appendix.}. For the following analysis, we refer to the three peaks in each emission line as the ``red", ``blue" or ``central" peak. We fit each of the three peaks in all emission lines with their own Gaussian. We also assume that all correspondingly shifted components individually abide by the constraints described earlier in Section \ref{sec:emission} (e.g., the separation between the two red peaks, the two blue peaks and the two central peaks of the \NII \: doublet are all constrained to be the rest-frame wavelength separation).

We first fit the \SII \: doublet, fixing the relative wavelength separation between the two red peaks, the two blue peaks and the two central peaks. This fitting results in red/blue peaks offset from the central component by $+$145 km s$^{-1}$ and $-$144 km s$^{-1}$. We use the line profiles for the red, blue and central peaks from this fit as a model for the corresponding peaks in the H$\alpha$ and \NII \: lines and constrain their widths in velocity space, as in the previous fits (see Section \ref{sec:emission}). We again allow for a broad H$\alpha$ component if the {reduced} $\chi^2$ is reduced by at least 20\%. We find that the resulting best-fitting model includes three narrow peaks for the \NII \: doublet and the H$\alpha$ line and one broad H$\alpha$ component\footnote{This galaxy had a ``hint" of broad \ha \: in the GAMA spectra, meaning that there was some asymmetry suggesting that the line would be best fit with multiple components.}. The red and blue peaks {for both H$\alpha$ and \NII \:} lie $-$138 km s$^{-1}$ and $+$146 km s$^{-1}$ offset from the central peak. We measure the broad H$\alpha$ luminosity and FWHM as L(H$\alpha$)$_b=10^{40.6}$ erg s$^{-1}$ and FWHM(H$\alpha$)$_b$=1609 km s$^{-1}$, giving a BH mass of M$_\mathrm{BH}=10^{6.3}$ \msun.

We follow a similar process for the H$\beta$ line as for the H$\alpha$, by constraining the Gaussian widths in velocity space based on the respective widths of the various peaks of the \SII \: doublet and allowing a broad component if the {reduced} $\chi^2$ value is reduced by 20\% or more. {The best fit model for this emission line includes just three narrow peaks. The red/blue peaks are offset by $+$155 km s$^{-1}$ and $-$151 km s$^{-1}$ from the central peak.}

Lastly, we fit the \OIII $\lambda$5007 line. This line does not exhibit three obvious visible peaks. {However, since the other lines do show three peaks, we test multiple methods of fitting the \OIII\ emission line and compare the resulting {reduced} $\chi^2$ values.} The models we test are:

\begin{enumerate}
    \item Model 1: a single Gaussian component model 
    \item Model 2: a two component model (a central narrow peak + a broad component)
    \item Model 3: a four component model (three red/central/blue narrow peaks + a broad component) 
\end{enumerate}

{Models 1 and 2 are the same as the models we consider in Section \ref{sec:emission} for fitting a single-peaked \OIII \: emission line coming from a central AGN.} Model 2 is clearly a better fit than Model 1, reducing the {reduced} $\chi^2$ value significantly (from 1.531 to 0.927), with the broad component offset from the narrow component by $-$30 km s$^{-1}$ and having a velocity dispersion of 182 km s$^{-1}$.

Given the triple-peaked nature of the other emission lines, we consider Model 3 to test whether the \OIII\ line may be also be fit with three peaks. However, the central peak clearly dominates the line profile and is due to emission from a central AGN (as evidenced by line ratios in both the MagE, see section \ref{sec:emission_line_ratios}, and GAMA spectra). Nevertheless, we aim to determine whether or not there may be additional, presumably weak, red and blue peaks in the \OIII\ line and constrain their properties.
We first fit Model 3 without constraints, but the fitting process always resulted in one of the peaks having a flux of $\sim$ zero.  Next, we force the profiles of the red/blue peaks of \OIII\: to emulate the other emission lines by constraining the offsets, widths and flux ratios of the red and blue peaks to be approximately within the range seen for the other emission lines that exhibit obvious triple peaks. Modeling the \OIII\ emission line in this way results in red/blue peaks that are offset by $-$131 km s$^{-1}$ and $+$156 km s$^{-1}$, respectively, and they are both extremely sub-dominant with respect to both the narrow and broad components of the central peak. Moreover, the width of the \OIII\ central peak is much larger than the central peak in the other emission lines, with a dispersion of 80 km s$^{-1}$ compared to $\sim$50 km s$^{-1}$ for the other lines. The broad component is offset from the central peak by $-$60 km s$^{-1}$ and has a velocity dispersion of 203 km s$^{-1}$. The {reduced} $\chi^2$ value for this model is 0.986, comparable to that of Model 2 (see Figure \ref{fig:oiii_fits}). 

Emission line fluxes for this object are reported in Table \ref{tab:fluxes} and the fits are shown in Figure \ref{fig:spec_all_2}. We note that the three peaks observed in all of the emission lines (except \OIII) are remarkably symmetric. The red/blue peaks are each offset by $\sim$140 km s$^{-1}$ and all three peaks have similar velocity dispersions ranging from 44-61 km s$^{-1}$ with an average of 49 km s$^{-1}$. 
{We again use the width of the \NII \: emission line as a proxy for the galaxy stellar velocity dispersion in the same way as for the other galaxies lacking strong absorption lines, as described in Section \ref{sec:nii_sigmas}. Since this galaxy is triple-peaked, however, we re-fit the \SII \: and \NII \: emission lines with a single gaussian for each line in the doublet to obtain an estimate of the total \NII \: emission line width. We find a value of 157 km s$^{-1}$ for this galaxy.} We will discuss possible physical interpretations for the triple-peaked lines in Section \ref{sec:triple_peaked_interpretation}.

\begin{deluxetable*}{ccccccc}
\tablecaption{BH Masses and Velocity Dispersions}
\tablewidth{0pt}
\tablehead{
\colhead{ID} & \colhead{log L$\mathrm{{(H\alpha)_b}}$} & \colhead{FWHM$\mathrm{{(H\alpha)_b}}$} & \colhead{log M$_\mathrm{BH}$} & \colhead{$\sigma_{\NII}$} & \colhead{$\sigma_{*,\NII}$} & \colhead{\sigstar} \\
\colhead{} & \colhead{[erg s$^{-1}$]} & \colhead{[km s$^{-1}$]} & \colhead{[M$_\odot$]} & \colhead{[km s$^{-1}$]} & \colhead{[km s$^{-1}$]} & \colhead{[km s$^{-1}$]} \\
\colhead{(1)} & \colhead{(2)} & \colhead{(3)} & \colhead{(4)} & \colhead{(5)} & \colhead{(6)} & \colhead{(7)} 
}
\startdata
GAMA 2258819 & 40.9 & 1252 $\pm$ 135 & 6.2 & 50 $\pm$ 9 & 50$^{+19}_{-16}$ &  \nodata  \\
GAMA 2123678\tablenotemark{\footnotesize{a}}  & \nodata & \nodata & \nodata & 25 $\pm$ 4 & 25$^{+7}_{-6}$ &  36 $\pm$ 20  \\
5227891  & 40.6 & 1609 $\pm$ 159 & 6.3 & 157 $\pm$ 21 & 157$^{+44}_{-37}$ &  \nodata  \\ 
GAMA 5240292 & 42.2 & 2030 $\pm$ 25 & 7.3 & 80 $\pm$ 7 & 80$^{+14}_{-12}$ &  \nodata  \\
GAMA 5275222 & 42.6 & 1167 $\pm$ 12 & 7.0 & 76 $\pm$ 5 & 76$^{+11}_{-9}$ &  \nodata  \\
SDSS J1605+0850\tablenotemark{\footnotesize{a}} & \nodata & \nodata & \nodata & 37 $\pm$ 5 & 37$^{+10}_{-8}$ &  46 $\pm$ 10  \\
\enddata
\tablecomments{
Column 1: Identification number in GAMA/SDSS.
Column 2: Broad H$\alpha$ luminosity. Uncertainties on this are $\sim$3\%.
Column 3: Broad H$\alpha$ FWHM.
Column 4: Virial BH mass estimated using fits to the broad H$\alpha$ emission line. Uncertainties on these masses are $\sim$0.5 dex.
Column 5: Gas velocity dispersion measured from the \NII \: emission line width. These values are used as a proxy for the stellar velocity dispersion in Figure \ref{fig:mbh_sigma_relation}.
Column 6: Stellar velocity dispersion measured from the gas velocity dispersion. Uncertainties from both the line widths and from the relation between $\sigma_{\NII}$ and \sigstar \: \citep{Barth2008} are added in quadrature.
Column 7: Stellar velocity dispersion measured from fits to the stellar absorption line spectra using \texttt{ppxf}. For ID 2123678 we report the $\sigma_*$ calculated from fitting the \mgb \: complex while for ID J1605+0850 we report the $\sigma_*$ calculated from fitting the \CaII \: triplet. {A three-dot ellipsis in this column indicates that the galaxy did not have strong absorption lines present in its spectrum and, therefore, we could not measure the stellar velocity dispersion using \texttt{ppxf}.}
\tablenotetext{a}{Broad H$\alpha$ was not detected for these objects, so we are not able to calculate the broad H$\alpha$ luminosity/FWHM or BH mass.
}
}
\label{tab:results}
\end{deluxetable*}


\section{Scaling Relations}
\label{sec:scaling_relations}

Using our virial BH masses and velocity dispersion measurements, we investigate how our sample compares to two different canonical scaling relations and other samples of galaxies. We consider the \mbh $-$ \sigstar \: and the \mbh $-$ \mgal \: relations.

\subsection{The M$_{BH}$ $-$ \sigstar \: Relation}
\label{sec:relation}

We place our galaxies on the \mbh $-$ \sigstar \: relation in Figure \ref{fig:mbh_sigma_relation}. For comparison, we also include the sample of classical bulges and elliptical galaxies with dynamical BH mass measurements from \citet{Kormendy2013} and the samples of low-mass AGNs from \citet{Xiao2011} and \citet{Baldassare2020}. \citet{Xiao2011} investigated a sample of 76 Seyfert 1 galaxies with low-mass BHs from SDSS {that were originally presented in \citet{Greene2007}}. \citet{Baldassare2020} looked at a sample of eight dwarf galaxies that were also from SDSS and identified as broad-line AGNs in \citet{Reines2013}. {Additionally, in their work, \citet{Baldassare2020} considers seven other objects with previously measured BH masses/velocity dispersions, including NGC 4395 \citep{Filippenko1989, Filippenko2003}, Pox 52 \citep{Barth2004, Thornton2008}, RGG 118 \citep{Baldassare2015, Baldassare2017}, RGG 119 \citep{Baldassare2016} and three galaxies with dynamical BH mass measurements, M 32 \citep{vandebosch2010}, NGC 5206 and NGC 205 \citep{Nguyen2018, Nguyen2019}.} In both of these works, they find that the galaxies are generally consistent with the relation defined by the higher-mass, inactive counterparts. {Our work generally agrees with both of these findings. Most of our galaxies lie within the $3\sigma$-scatter for the canonical relation defined by \citet{Kormendy2013}. The only galaxy lying below the relation is the triple-peaked emission line galaxy, whose BH and \sigstar \: measurements were complicated by the triple-peaked emission lines.} 

\citet{Volonteri2009} show that the initial BH seed mass should be reflected in the low-mass regime of this relation. Although the seeds do not initially land on the canonical relation, galaxies tend to migrate onto the relation through merger and accretion events, with higher mass galaxies tending to migrate faster. {They find that a heavy seeding model results in a population of galaxies with over-massive BHs at low-mass and, therefore, a broadening of the relation. On the other hand, they find that a light seeding model tends to manifest in a downturn of the relation. \citet{Pacucci2018} estimate that this downturn should occur at \sigstar$\approx$65 km s$^{-1}$. From our work, we cannot distinguish between these seeding models. We do not observe any galaxies with significantly over-massive BHs especially at low \sigstar \: that would be indicative of a heavy seeding model.

\subsection{The M$_{BH}$ $-$ M$_{*}$ Relation}

{Originally, BH mass was found to correlate with bulge mass \citep[e.g.,][]{Haring2004, Kormendy2013, McConnell2013}. However, comparing BH mass to total galaxy stellar mass has recently become a useful tool since extracting bulge masses can be a complicated process and is not feasible in high-redshift galaxies. Moreover, multiple works have shown that galaxies of different types than those originally used to create the canonical bulge mass relation defined by \citet[][i.e. large, local, elliptical/classical bulge galaxies]{Kormendy2013} tend to deviate from it, including dwarfs, late-type galaxies, high-redshift galaxies and AGNs \citep{Wandel1999, Greene2008, Greene2010, Hu2008, Gadotti2009, Kormendy2011, Sani2011, Graham2014, Reines2015, Sturm2024}. }

We show our sample on the BH mass versus galaxy total stellar mass (\mgal) scaling relation using masses and uncertainties reported in tables \ref{tab:sample} and \ref{tab:results}. We compare our sample to the BH-stellar mass relations derived in \citet{Reines2015} both for local broad-line AGNs (predominantly in galaxies with disks) and for quiescent, early-type galaxies in Figure \ref{fig:mbh_mgal}.

{Our galaxies are generally consistent with the canonical scaling relations. Three galaxies (IDs 2258819, 5275222 and 5240292) are consistent within 3$\sigma$ of the intrinsic scatter of both relations. Our triple-peaked galaxy (GAMA 5227891) falls below the relation for early-type galaxies, but is consistent with the local AGNs from \citet{Reines2015}.}

\begin{figure}
    \centering
    \includegraphics[width=8cm]{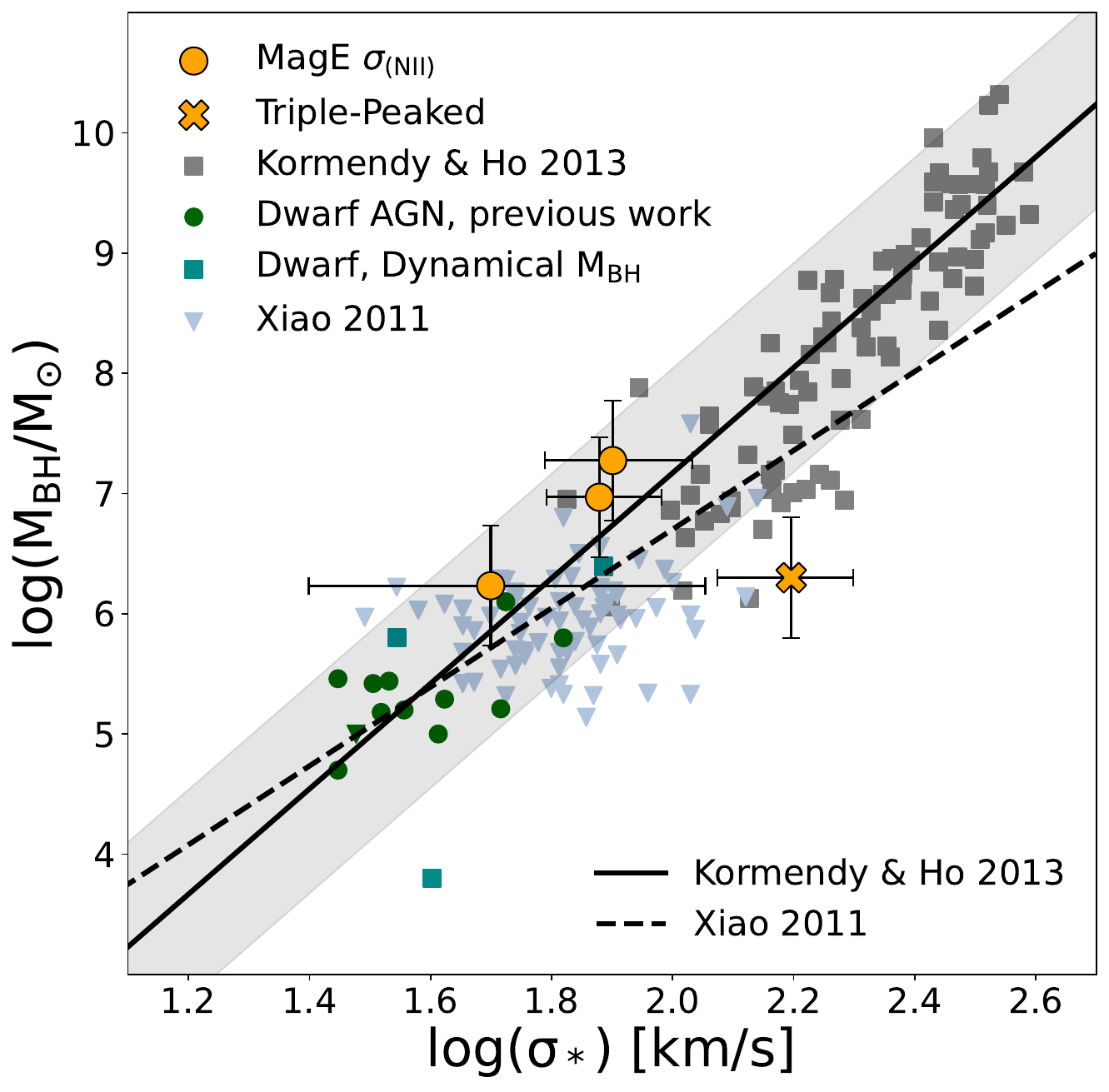}
    \caption{The \mbh $-$ \sigstar \: relation. Our galaxies are shown in orange with $\sigma$ derived using \NII. Vertical error bars represent the $\sim$0.5 dex uncertainty involved in using virial BH mass estimates and horizontal error bars incorporate both the uncertainty in the width of the \NII \: lines and the 0.15 dex uncertainty in the $\sigma_*-\sigma_{\mathrm{g}}$. {For comparison, we also show the sample of elliptical and classical bulge galaxies from \citet{Kormendy2013} and the sample of low-mass AGNs from \citet{Xiao2011}. We also show dwarf AGN and dwarf galaxies with dynamical BH masses, which includes the sample from \citet{Baldassare2020} as well as their additional objects from the literature described in Section \ref{sec:relation}. } The solid line shows the best fit relation derived from the elliptical and classical bulges from \citet{Kormendy2013} and the shaded region indicates the 3$\sigma$ intrinsic scatter. We also show the relation found by \citet{Xiao2011} for their sample of low-mass AGNs (dashed line).}
    \label{fig:mbh_sigma_relation}
\end{figure}

\begin{figure}
    \centering
    \includegraphics[width=8cm]{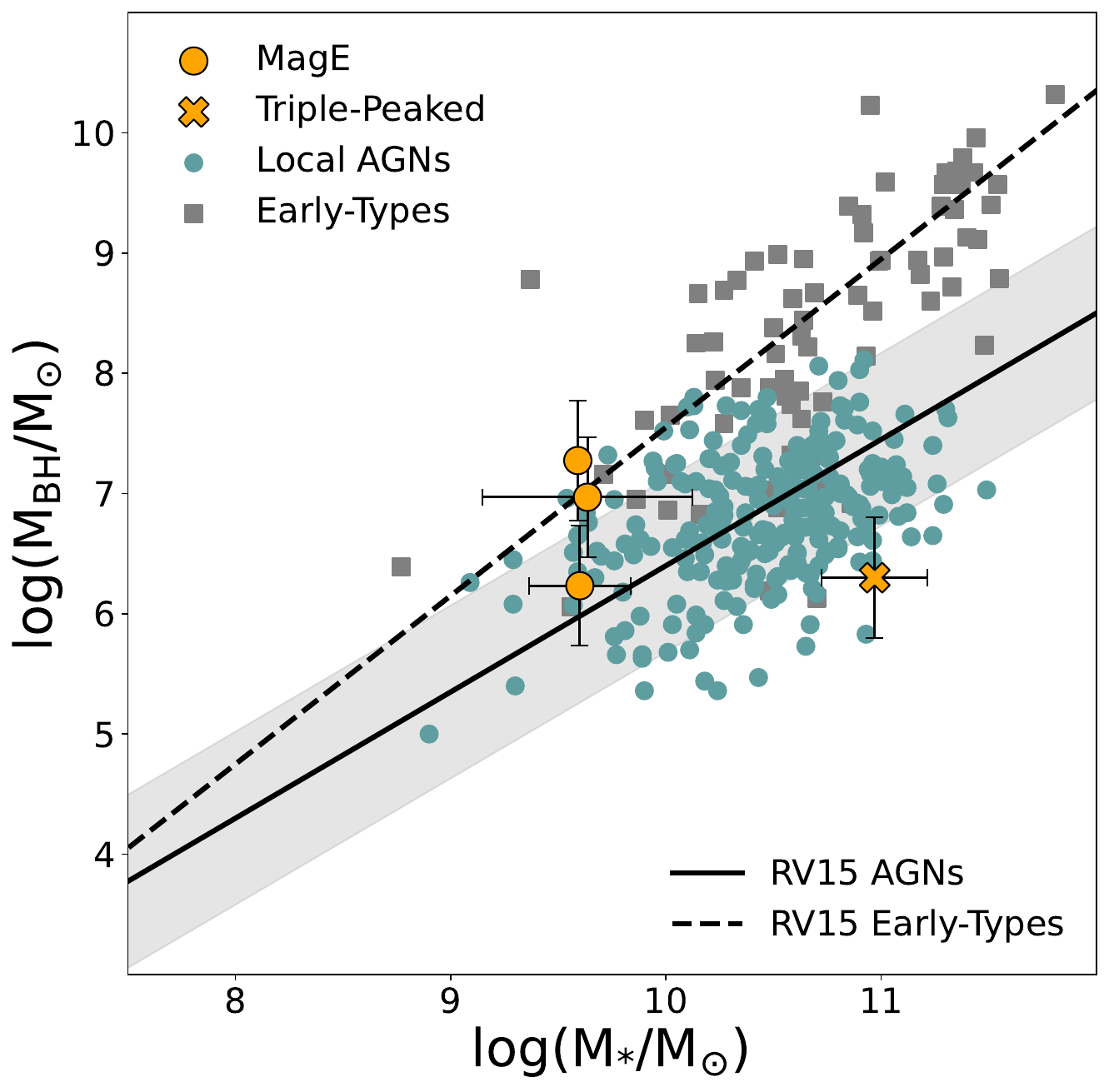}
    \caption{{The BH mass versus total galaxy stellar mass (\mbh $-$ \mgal) relation.} Our galaxies are shown in orange. Vertical error bars represent the $\sim$0.5 dex uncertainty involved in using virial BH estimates and horizontal error bars are the reported uncertainties in stellar mass from \citet{Taylor2011} and \citet{Bellstedt2020}. {For comparison, we show} {the sample of local AGNs presented in \citet{Reines2015} and the sample of early-type galaxies (elliptical and S/S0 with classical bulges) that have dynamically-detected BHs from \citet{Kormendy2013} with total stellar masses calculated in \citet{Reines2015}.} We also show the BH-stellar mass relations derived by \citet{Reines2015} for the AGNs (solid line) and for the early-types (dashed line).}
    \label{fig:mbh_mgal}
\end{figure}

\begin{figure*}
    \centering
    \includegraphics[width=18cm]{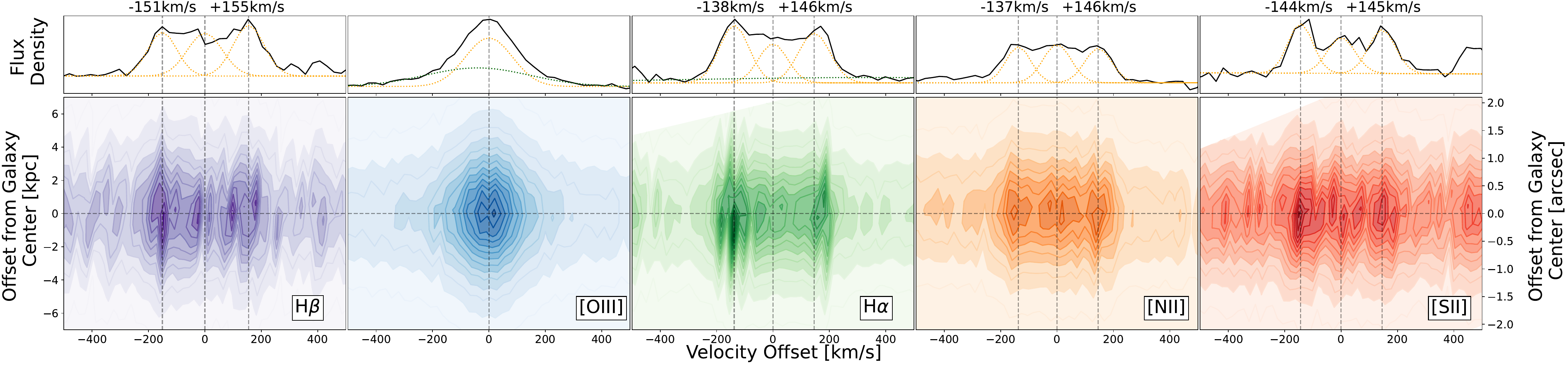}
    \caption{2-D (bottom) and 1-D (top) spectra for galaxy GAMA 5227891, which exhibits triple-peaked emission lines. In the top panel, orange lines show the best fit narrow line components and green lines show the best fit broad components. In the bottom panel, the horizontal grey dotted line shows the center of the galaxy within the slit. The vertical grey dotted lines show the center points of the red and blue shifted line components, and their offset from the center line is given at the top of each line. We give the offset from the galaxy center both in kpc and arcseconds.}
    \label{fig:spec2d_5227891}
\end{figure*}

\subsection{Observational Bias}
\label{sec:obs_bias}

We note that our sample is small and likely a result of observational bias towards higher luminosity objects. The four objects we placed on the \mbh $-$ \sigstar \: relation are at redshift $z\sim$0.2 and, although {most are low-mass} galaxies (with M$_*<5\times10^9$\msun), they host relatively large central BHs. {The three lowest-mass objects} fall above the \mbh $-$ \mgal \: relation defined by \citet{Reines2015} for local AGNs. Since smaller BHs have lower Eddington limits and are, therefore, less likely to be accreting at the levels needed to be observed at such large distances \citep{Schulze2010}, it is possible that we are simply missing a population of dwarf galaxies occupying regions of the \mbh \: versus \sigstar \: or \mgal \: relation that are inconsistent with the canonical relations. For example, since we are effectively observationally limited in BH mass, there could be a downturn in the relation (i.e., galaxies with \mbh \: $<$ 10$^{4.5}$ \msun \: and \sigstar $\sim$ 10$^{1.6}$) that we are not able to observe. We are also limited by the resolution of our spectra ($\sim$ 20 km s$^{-1}$), meaning that we are not able to detect galaxies with smaller velocity dispersions.

\section{Investigating the Triple-Peaked Emission Line Galaxy}
\label{sec:triple_peaked_interpretation}

{Here we return to the triple-peaked emission line galaxy to further analyze its spectrum and discuss possible physical interpretations. In Section \ref{sec:triple_peaked}, we fit the \ha, \NII, \SII \: and \hb \: emission lines each with three narrow gaussian peaks. For all of these emission lines, the red/blue peaks have similar velocity dispersions and offsets from the central peak. Since the \OIII \: line did not have clear triple peaks, we tested three different model fits, ultimately finding that a two component model (one broad + one narrow peak) was the best fit. However, using a four component model (a central dominant peak with a narrow + broad component, as well as weak narrow red and blue peaks) also resulted in a similar {reduced} $\chi^2$ value. We will consider both of these models in the following discussion of the possible physical origins of the emission in this galaxy. } 

\subsection{Spatial Distribution of Emission}

In Figure \ref{fig:spec2d_5227891}, we show the 2-D spectrum around the H$\beta$, \OIII $\lambda$5007, H$\alpha$, \NII $\lambda$6585 and \SII $\lambda$6716 emission lines. This shows the spatial position of the gas (offset from the galaxy center) versus velocity offset (relative to the central peak of the emission line fit). It is immediately apparent that both the spatial and velocity distribution of the \OIII\ emission is markedly different from the other emission lines. The \OIII\ emission is clearly dominated by a central source (i.e., the AGN). In contrast, the other lines have significant emission offset in both velocity and position. There is not a consistent decrease in emission at higher velocities/radii (as seen in the \OIII\ line), indicating that the primary emission source for these lines is likely further out in the galaxy than the primary source of the \OIII \: emission.

\begin{figure*}
    \centering
    \includegraphics[width=16cm]{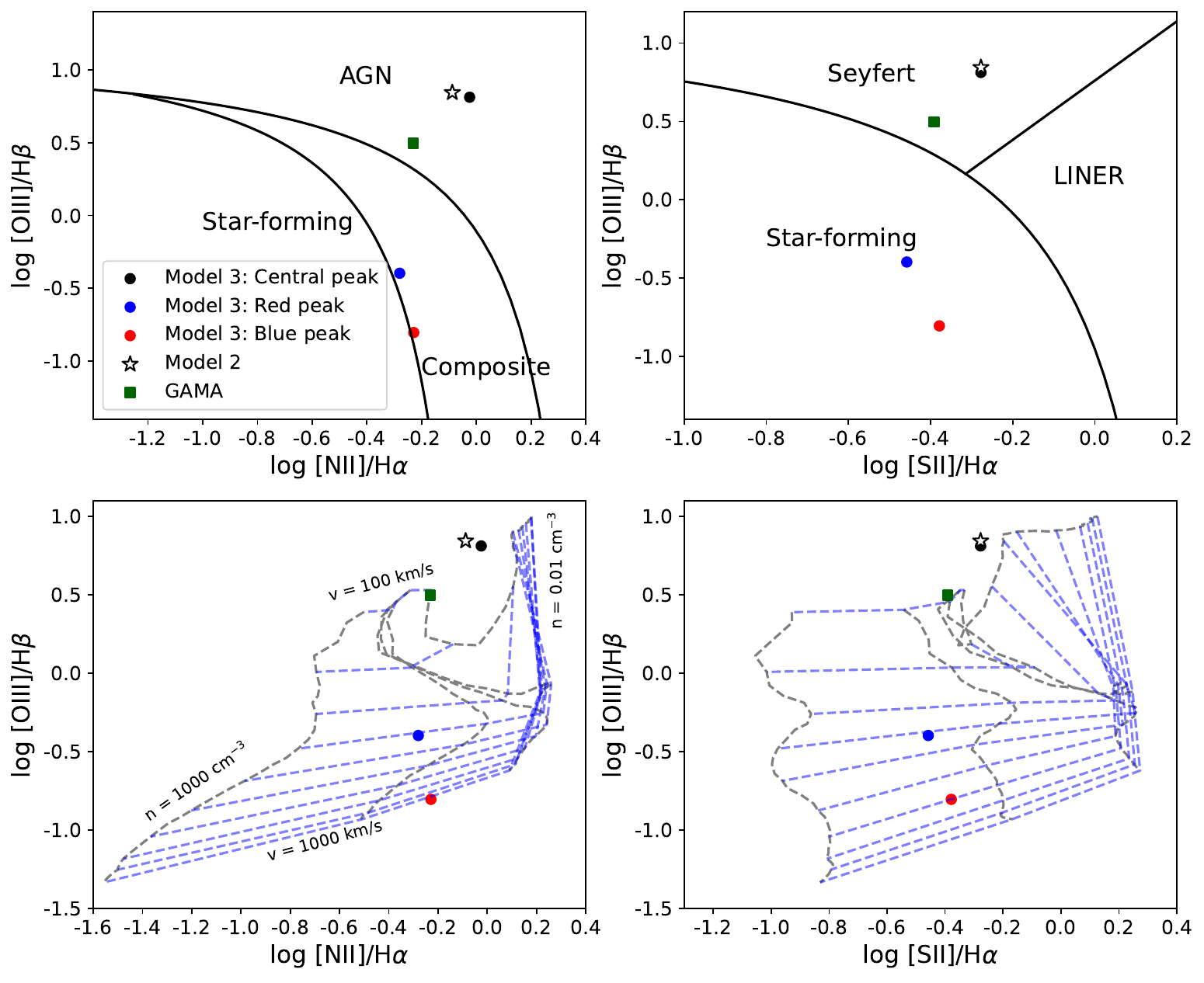}
    \caption{{Top:} Narrow emission line ratio diagnostic diagrams for various fits to the \OIII \: emission line. Based on the GAMA spectrum, the galaxy lies in the AGN/Seyfert region (shown as the green square). Using Model 2, the galaxy lies in the AGN/Seyfert region (shown as the black star). For Model 3, the central peak lies in the Seyfert region of both the \OIII/H$\beta$ versus \NII/H$\alpha$ and \OIII/H$\beta$ versus \SII/H$\alpha$ diagrams (shown as the black dot), and is in almost the same position as for Model 2. On the other hand, both the red and blue peaks lie in the composite/star-forming regions (shown as the red/blue dots).
    {{Bottom:} Shock excitation models from \citet{Allen2008} with fixed transverse magnetic field of b = 1 $\mu$G, velocities ranging from 100-1000 km s$^{-1}$ and electron densities ranging from 0.01-1000 cm$^{-3}$. The red and blue peaks for Model 3 are also consistent with ionization due to shocks with gas density n$\sim$100 cm$^{-3}$ and at velocities of $\sim$400-1000 km s$^{-1}$.}}
    \label{fig:bpt}
\end{figure*}

\subsection{Ionization Sources and Gas Density}
\label{sec:emission_line_ratios}

We investigate the ionization source(s) of the gas in {the triple-peaked system} by studying the emission line ratios of the various peaks. Following broadly from the idea that ionization from AGNs can produce more high energy photons than stellar processes, we can distinguish AGN emission by looking for higher ratios of collisionally-excited, forbidden emission lines (such as \NII $\lambda$6583, \SII $\lambda\lambda$6716,1731 and \OIII$\lambda$5007) to a line such as H$\alpha$ or H$\beta$. For the \OIII/H$\beta$ versus \NII/H$\alpha$ diagram \citep[BPT;][]{Baldwin1981}, this separation has been defined in \citet{Kewley2006} by combining a distinction between purely star-forming/\HII \: galaxies and composite galaxies from \citet{Kauffmann2003c} with an extreme starburst limit from \citet{Kewley2001}, above which (theoretically) only AGNs could produce the observed emission. \citet{Kewley2006} also outlines a classification scheme for the \OIII/H$\beta$ versus \SII/H$\alpha$ diagram, using the theoretical extreme starburst limit from \citet{Kewley2001} again to separate AGNs from star-forming galaxies in addition to a Seyfert-LINER separation line.

For this analysis, we consider both Model 2 and Model 3 for the \OIII \: emission line. The \OIII\ flux for Model 2 includes both the narrow and broad component.  The central peak for Model 3 includes the central narrow component plus the broad component. These give nearly identical results on the diagnostic diagrams (since any possible red/blue peaks are quite weak) and demonstrate that AGN emission dominates (Figure \ref{fig:bpt}). For comparison, we also show the results of \citet{Salehirad2022} using the GAMA spectrum. The GAMA spectrum is at lower resolution than the MagE spectrum we use in this work and the three peaks were not resolved in GAMA. Therefore, the GAMA spectrum shows the dominant emission source in the galaxy (i.e., an AGN).

For Model 3, the red and blue peaks lie in the composite and star-forming regions in the \NII/H$\alpha$ and \SII/H$\alpha$ diagrams, respectively. Therefore, if these red/blue peaks in the \OIII\ line do indeed exist, the line ratios suggest a star-formation origin (or shock ionization, see below). These findings are consistent with the fact that we observe significant emission (e.g., in H$\alpha$) that is red/blue-shifted and originating from outer regions in the galaxy (see Figure \ref{fig:spec2d_5227891}).

{Motivated by the AGN outflow observed in the \OIII \: emission line and the BPT location of the red/blue peaks of Model 3, we also compare our emission line ratios to models of shock ionization.  We use shock models from \citet{Allen2008} with a transverse magnetic field of b = 1 $\mu$G and varying velocities (v = 100-1000 km s$^{-1}$) and electron densities (n = 0.01-1000 cm$^{-3}$). We show these models with our emission line ratios overlaid in the bottom panels of Figure \ref{fig:bpt}. 
The red and blue peaks in Model 3 are consistent with shock models for an electron density of n$\sim$100 cm$^{-3}$ and shock velocities of $\sim$400-1000 km s$^{-1}$.
These shock velocities are comparable to the outflow velocity of $\sim$500 km s$^{-1}$\footnote{Following \citet{salehirad2024} and \citet{ZakamskaGreene2014}, we define the outflow velocity as W$_{80}$, which is the line with including 80\% of the flux. For a gaussian, W$_{80}$=1.09FWHM.} derived from the broad component of the \OIII \: emission line (using either Model 2 or 3). We also examined shock + precursor models and found that the observed line ratios are not consistent with those models.

Finally, we estimate the electron density ($n_e$) of each peak using the flux ratio of the \SII \: doublet. We follow \citet{Sanders2016} and compute $n_e = (cR -ab)/(a-R)$, with $R$ = \SII$\lambda$6716/\SII$\lambda$6731, $a=0.4315, b=2107$ and $c=627.1$. We note that $n_e$ saturates at a value of $R \sim 1.45$, however this can still give us a good estimate of the relative electron densities for each peak. The central peak has the lowest value of $R=1.06$ and, therefore, the highest electron density of $\sim$10$^{2.6}$ cm$^{-3}$, which is consistent with typical values for the NLR around low-mass BHs \citep[$10^2-10^3$ cm$^{-3}$, ][]{Ludwig2012}. The red peak has a slightly higher value of $R=1.25$ and lower electron density of $10^{2.2}$ (consistent with the shock models described above), while the blue peak has a much lower electron density, saturating the flux ratio with $R=1.78$. The lower electron densities in the blue/red peaks compared to the central peak is consistent with the scenario presented above in which the central peak is dominated by the AGN, while the blue/red peaks may have star-formation origin. This is qualitatively similar to the results in \citet{Zhang2024}, who study a sample of Type-1 AGNs, Type-2 AGNs and \HII \: galaxies, and find that the \HII \: galaxies have mean $R \sim 1.4$, while the AGNs tend to have lower ratios ($R \sim 1.13$ for Type-1 and $R \sim 1.21$ for Type-2), and thus higher electron densities.

\subsection{Physical Interpretation}
\label{sec:physical_interp}

Here we discuss possible physical interpretations for the triple-peaked emission lines in the galaxy GAMA 5227891. Multi-peaked emission lines are not unprecedented, however, it can be difficult to distinguish between the various possible physical origins including a rotating gas disk (in the galaxy or NLR), bipolar outflows and multiple AGNs \citep[e.g.,][]{Muller-Sanchez2015}.

Samples of dual-AGN candidates have been produced by searching for double peaks (corresponding to distinct kinematics from the two AGNs) in the \OIII $\lambda\lambda$4960, 5007 emission lines \citep[][and references therein]{derosa2019}. These lines are typically used since they are expected to be the brightest optical emission lines resulting from AGN photoionization \citep{Rosario2010}. While we cannot definitively rule out the possibility that our galaxy hosts a system of three AGNs \citep[with each peak in the emission lines corresponding to a separate NLR, e.g.,][]{Benitez2018,Benitez2019}, we consider this highly unlikely for a number of reasons. First, even dual AGNs are quite rare, with \citet{Shen2011} suggesting that up to 90\% of the double-peaked \OIII \: lines result from jets or disks in the NLR, not two BHs. Moreover, the red/blue peaks of the \OIII\ line are very dim, if present at all (see \S\ref{sec:triple_peaked}).

While three AGNs seem exceedingly unlikely, the emission line ratios clearly indicate the presence of an AGN in the galaxy using the central peaks of the H$\alpha$, H$\beta$, \NII, and \SII\ emission lines regardless of the model used for the \OIII\ line (Figure \ref{fig:bpt}). Given the justified assumption of a single AGN in the galaxy and that the \OIII\ line profile is dominated by a central peak with a narrow + {\it broad} component, we also infer the presence of an outflow associated with the central AGN \citep[e.g.,][]{salehirad2024}.

We now turn to possible origins for the red/blue peaks robustly detected in the other emission lines (e.g., H$\alpha$, H$\beta$, \NII, \SII). The red/blue peaks may be produced by ionized gas that is in some sort of kinematic structure such as a rotating ring or disk.  A bipolar outflow is also a possibility, although the red and blue peaks in the aforementioned emission lines have a relatively symmetric appearance and flux ratios consistent with unity (Table \ref{tab:fluxes}), favoring a disk scenario \citep[e.g.,][]{Smith2012}. The spatial extent of the aforementioned emission lines seen in Figure \ref{fig:spec2d_5227891} provide supporting evidence for an extended structure distinct from the nucleus. Given the relatively weak \OIII\  emission at high velocity and spatial offsets and the associated line ratios, the gas in this structure is likely ionized predominantly by star formation {or shocks}. 

Shocks produced by AGN outflows have been observed to trigger star formation in the dwarf galaxy Henize 2-10 \citep{Schutte2022} and we consider such a scenario a possibility here, albeit with some significant differences. There is unambiguous evidence for a central AGN in GAMA 5227891 based on optical line ratios and the robust detection of a broad component in the \OIII\ emission line is indicative of an outflow from the AGN. While speculative, it is possible that the AGN-driven outflow takes the form of non-collimated ionized winds that are impacting the interstellar medium further out in the galaxy, resulting in a ring/disk structure of shocked gas. While such a scenario is consistent with the observations described above, additional observations are needed to test this hypothesis.


\section{Conclusions}

Investigating the behavior of scaling relations in the low-mass regime is crucial in the context of understanding BH/galaxy formation and coevolution. Relating BH mass to galaxy velocity dispersion and stellar mass can provide insight to both the primary seeding mechanism at play in the early universe as well as important growth processes for both the BHs and their host galaxies. 

In this work, we present new high-resolution MagE spectra of six galaxies previously identified as hosts to AGN \citep{Reines2013, Salehirad2022} in an effort to help populate the low-mass end of BH scaling relations. We fit both the galaxy absorption line spectra and the AGN emission line spectra, allowing us to make velocity dispersion measurements and estimate BH masses based on broad H$\alpha$ emission and standard virial techniques. We place our galaxies on two scaling relations, comparing BH mass to both stellar velocity dispersion and total galaxy stellar mass. We also identify a triple-peaked emission line galaxy hosting an AGN.
Our main findings are as follows:

\begin{enumerate}

    \item {For the low-mass galaxies in our sample (\mgal $<5\times10^9$ \msun), we measure velocity dispersions in the range $\sigma \sim 25-80$ km s$^{-1}$. For the more massive, triple-peaked emission line galaxy, we measure a velocity dispersion of 157 km s$^{-1}$.} Four of the galaxies have detectable broad H$\alpha$ emission with corresponding BH mass estimates of $M_{\rm BH} \sim 10^{6.2-7.3}~M_\odot$.

    \item Our sample is generally consistent with the canonical \mbh $-$ \sigstar\ relation at the low-mass end.

    \item Our sample exhibits more scatter around the \mbh $-$ \mgal\ relation, but the small sample size prevents us from making any strong conclusions regarding BH seeding models.

    \item Our MagE spectrum of one galaxy {(\mgal = $10^{10.97}$ \msun)} reveals triple-peaked emission lines in H$\alpha$, H$\beta$, \NII, and \SII\ that were not resolved in GAMA. However, the \OIII\ line profile is dominated by a single peak with an additional broad component, likely signifying an AGN outflow. The red/blue peaks detected in the other lines can be attributed to an extended ring/disk primarily ionized by star formation or shocks from the AGN.  
    
\end{enumerate}

\vspace{5mm}

A.E.R. gratefully acknowledges support for this work provided by NSF through CAREER award 2235277.


\appendix
\label{sec:appendix}

We ensure that the emission lines in galaxy GAMA 5227891 are truly triple-peaked and not just a product of our data reduction by visually examining the 2-D spectra before performing any data reduction. If the sky modeling/subtraction is done incorrectly in the reduction process, it could potentially cause artificial triple peaks in the emission lines. We show both the raw and post-reduction 2-D and 1-D spectra around each emission line in Figure \ref{fig:reduction_check}. Since the multiple peaks are discernible even before we perform any sky-subtraction, we know that the data reduction pipeline is not the source of the triple-peaks and instead there is a physical origin for this observation.

\begin{figure*}
    \centering
    \includegraphics[width=18cm]{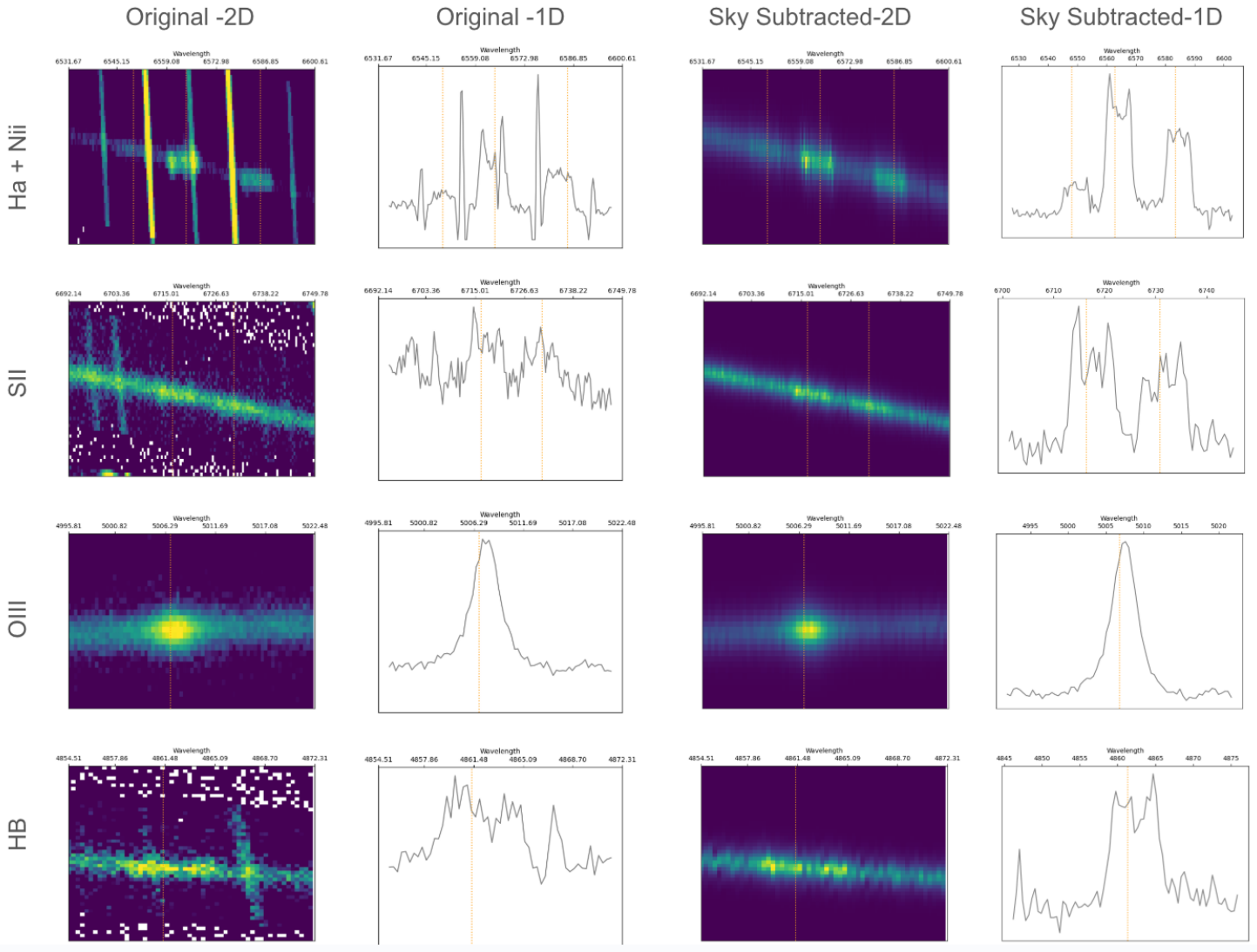}
    \caption{The raw 2-D (first column) and 1-D (second column) spectra and the sky-subtracted 2-D (third column) and 1-D (fourth column) spectra resulting from the reduction pipeline for our triple-peaked galaxy GAMA 5227891. The triple peaks are visible both in the raw data and in the sky-subtracted images, indicating that the triple peaks are real and not introduced as a result of the reduction process.}
    \label{fig:reduction_check}
\end{figure*}


\bibliography{mybib}

\end{document}